\documentclass[onecolumn,preprint,aps,superscriptaddress,amsmath,secnumarabic,pdflatex,
amssymb,prl]{revtex4}
\usepackage{bm}
\usepackage{amsmath,amsfonts,amssymb}
\usepackage{epstopdf}
\usepackage{color}

\usepackage{graphicx}

\begin{document}
\title{High Efficiency Gamma-Ray Flash Generation via Multiple Compton Scattering}

\author{Z. Gong}
\affiliation{State Key Laboratory of Nuclear Physics and Technology, and Key Laboratory of HEDP of the Ministry of Education, CAPT, Peking University, Beijing, China, 100871}
\author{R. H. Hu}
\affiliation{State Key Laboratory of Nuclear Physics and Technology, and Key Laboratory of HEDP of the Ministry of Education, CAPT, Peking University, Beijing, China, 100871}
\author{Y. R. Shou}
\affiliation{State Key Laboratory of Nuclear Physics and Technology, and Key Laboratory of HEDP of the Ministry of Education, CAPT, Peking University, Beijing, China, 100871}
\author{B. Qiao}
\affiliation{State Key Laboratory of Nuclear Physics and Technology, and Key Laboratory of HEDP of the Ministry of Education, CAPT, Peking University, Beijing, China, 100871}
\author{J. Chen}
\affiliation{State Key Laboratory of Nuclear Physics and Technology, and Key Laboratory of HEDP of the Ministry of Education, CAPT, Peking University, Beijing, China, 100871}
\author{X. He}
\affiliation{State Key Laboratory of Nuclear Physics and Technology, and Key Laboratory of HEDP of the Ministry of Education, CAPT, Peking University, Beijing, China, 100871}

\author{S. S. Bulanov}
\affiliation{Lawrence Berkeley National Laboratory, Berkeley, California 94720, USA}

\author{T. Zh. Esirkepov}
\affiliation{QuBS, Japan Atomic Energy Agency, Kizugawa, Kyoto, 619-0215, Japan}

\author{S. V. Bulanov}
\affiliation{QuBS, Japan Atomic Energy Agency, Kizugawa, Kyoto, 619-0215, Japan}
\affiliation{A. M. Prokhorov Institute of General Physics RAS, Moscow, 119991, Russia}

\author{X. Q. Yan\footnotetext{$\dagger$ x.yan@pku.edu.cn}}
 \email[]{x.yan@pku.edu.cn}
 \affiliation{State Key Laboratory of Nuclear Physics and Technology, and Key Laboratory of HEDP of the Ministry of Education, CAPT, Peking University, Beijing, China, 100871}
 \affiliation{Collaborative Innovation Center of Extreme Optics, Shanxi University, Taiyuan, Shanxi, China, 030006}

\date{\today}

\begin{abstract}
Gamma-ray flash generation in near critical density (NCD) target irradiated by four symmetrical colliding laser pulses is numerically investigated. With peak intensities about $10^{23}$ W/cm$^2$, the laser pulses boost electron energy through direct laser acceleration, while pushing them inward with the ponderomotive force. After backscattering with counter-propagating laser, the accelerated electron is trapped in the optical lattice or the electromagnetic standing waves (SW) created by the coherent overlapping of the laser pulses, and  emits gamma-ray photons in Multiple Compton Scattering regime, where electrons act as a medium transferring energy from the laser to gamma-rays. The energy conversion rate from laser pulses to gamma-ray can be as high as 50\%.
\end{abstract}

\maketitle

Gamma-rays are ubiquitous in the universe, from neutron stars, pulsars, supernova explosions, and regions around black holes \cite{klebesadel1973observations,woosley1993gamma}. They are also generated on earth \cite{sandmeier1972electromagnetic,reus1983catalog} by nuclear explosions, lightning, and radioactive decays. It's known that gamma-rays are widely used in radioactive tracer \cite{bergstrom2003positron} and treatment of malignant tumors \cite{baskar2012cancer}. Controllable intense gamma-ray sources are useful for laboratory astrophysics and space engineers to simulate the celestial processes and extreme environments \cite{aharonian2005new}. It is expected that powerful laser facilities like Extreme Light Infrastructure(ELI) \cite{korn}, which are aiming to deliver femtosecond pulses with intensities up to 10$^{24}$ W/cm$^2$, can provide new efficient regimes of gamma-ray generation. Thus there is a great demand for different gamma-ray sources both for applications and fundamental research.

Usually such sources use electron beams generated by conventional accelerators, but with the fast progress of Laser Plasma Acceleration (4.2 GeV electron beams are already reported from 9 cm long plasma \cite{Leemans_PRL_2014}) a new compact laser based design is being investigated. Laser driven gamma-ray sources can be divided into three types, depending on which processes of gamma production it is based on: (i) bremsstrahlung \cite{giulietti2008intense}, (ii) Compton scattering \cite{Phuoc_Nature_2012}, and (iii) Nonlinear Thomson Scattering radiation \cite{lobet2015ultrafast,liu2015quasimonoenergetic}. In these sources laser accelerated electrons collide with a counter-propagating laser pulse or oscillate in electric and magnetic fields, generated in plasma, radiating MeV photons by Compton scattering or Nonlinear Thomson Scattering  \cite{seipt2015narrowband,lobet2015ultrafast,liu2015quasimonoenergetic}.

In this paper, we present a numerical study of a novel laser plasma based gamma-ray source, by using four symmetrically imploding laser pulses and a near critical density (NCD) target. For laser intensity of $8.5\times10^{22}\,W/cm^2$, electrons experience multiple emissions of photons during the interaction, thus achieving Multiple Compton Scattering (MCS) regime. When laser intensities approaching $10^{23}$ W/cm$^2$, the nonlinear quantum electrodynamics (QED) effects begin to play a significant role in laser plasma interactions \cite{SVB-2015}. These effects manifest themselves through multi-photon Compton and Breit-Wheeler effects \cite{Ritus_1,Ritus_2,Ritus_3}, \textit{i.e.}, through either photon emission by an electron or positron, or electron-positron pair production by a high energy photon respectively. These processes are characterized in terms of two dimensionless parameters: $\chi_e^2 = -e^2(F^{\mu\nu}p_\nu)^2/m_e^6$ and $\chi_\gamma^2 = -e^2 (F^{\mu\nu}k'_\nu)^2/m_e^6$ \cite{Ritus_1}. Here $\hbar = c = 1$, $e$ and $m_e$ are electron charge and mass respectively, $F_{\mu\nu}$ is the EM field tensor, while $p_{\nu}$ and $k'_{\nu}$ denote the 4-momenta of electron or positron undergoing Compton process and photon undergoing Breit-Wheeler process. The probabilities of these processes depend strongly on $\chi_e$ and $\chi_\gamma$, reaching optimal values when $\chi_e\sim 1$ and $\chi_\gamma\sim 1$ \cite{Ritus_1}. The interaction of NCD plasma with four laser pulses at such intensities leads to the energy conversion rate from laser pulses to gamma-rays to be around 50\%. This is much higher than it is reported for the cases of the laser-NCD plasma interaction \cite{PhysRevLett.109.245006}, or laser-irradiated solids case \cite{ridgers2012dense,Nakamura_PRL_2012}, or two colliding laser pulses scheme\cite{nerush2011laser,Jirka_PRE_2016}. This scheme of gamma-ray generation may potentially become the most efficient gamma-ray source.

\begin{figure}[tbp]
\includegraphics[keepaspectratio=true,height=35mm]{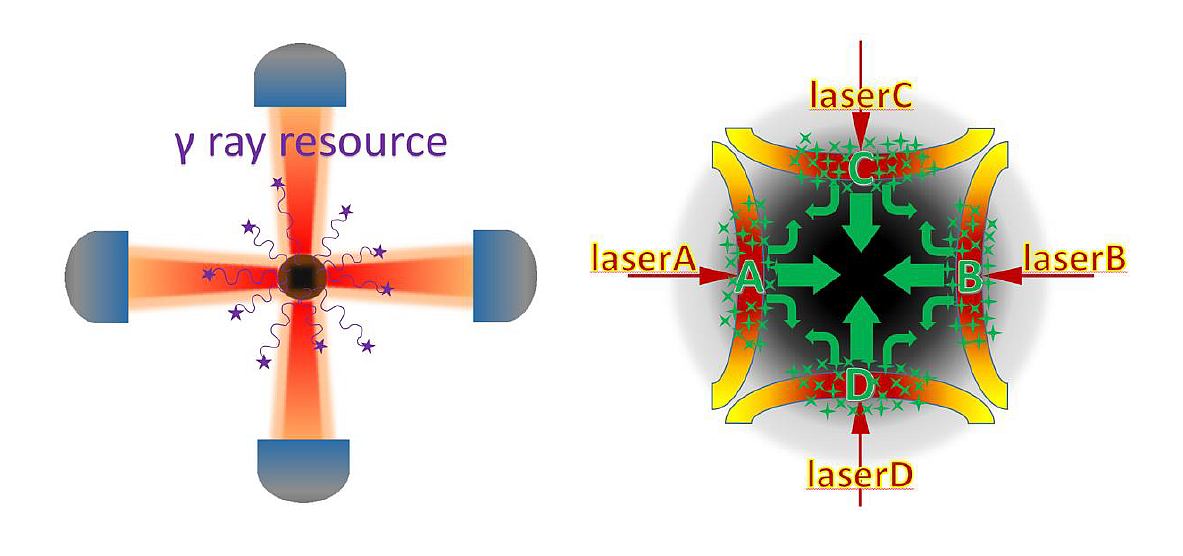}
\caption{
Schematic view of four-side irradiation. (a)  $8.5\times 10^{22}\,W/cm^2$ lasers irradiating a hydrogen target.
(b)The electrons are thrusted by driven laser and interact predominantly with the counter-propagating laser (for instance, electron A collides with laser B).
}
\label{fig1}
\end{figure}

For an electron interaction with a plane EM wave propagating along x-axis with $E=E(x-t)e_y$ and $B=E(x-t)e_z$ the parameters of interaction can be written in terms of EM field strength, normalized to the QED critical field, $E_S = m^2/e$ \cite{schwinger1951gauge} and either electron $\gamma$-factor or photon energy $\omega$:  $\chi_e=(E/E_S)(\gamma-p_x/mc)$ and $\chi_\gamma=(E/E_S)(\omega-k_x c)/mc^2$.  If an electron/positron or a photon co-propagates with the EM wave, then in the former case the parameter $\chi_e$ is reduced ($\chi_e\simeq(2\gamma)^{-1}(E/E_S)$) and in the later case the parameter $\chi_\gamma$ is equal to zero ($\chi_\gamma=0$). On the contrary, $\chi_e$ can be enhanced to approximately $2\gamma E/E_S$ \cite{chen2011radiation}, when electron interacts with a counter-propagating laser pulse. Therefore the head-on collision is an perfect scheme to enhance the production of $\gamma$ rays and it has been studied previously \cite{chen2013mev,kawase2008mev,bamber1999studies,Bulanov_PRA_2013}. Solid materials are usually used as the irradiated target experimentally, however, the production of $\gamma$ ray is limited because laser eventually reflects at relativistic critical-density (RCD) interface and its energy cannot further deplete in the plasma \cite{umstadter2003relativistic}. Therefore RCD plasma, with sufficient electrons to radiate, is an appropriate medium to transfer energy from laser to $\gamma$ rays. Compared with the two-side irradiation case\cite{nerush2011laser}, four-side symmetrical irradiation in Fig.\,\ref{fig1}(a) compresses the pellet more impeccably and moreover energetic electrons can be well confined in central region without substantial dispersing. Hence the configuration of four colliding pulses interacting with a RCD target has a positive effect for $\gamma$ photon production, where RCD can be achieved by a self-pileup from initial near critical density (NCD) material. We note that multiple colliding pulse scheme was first proposed to enhance the electron-positron pair production from vacuum by lowering the threshold laser intensity \cite{Bulanov_PRL_2010} and then utilized to study the effective particle trapping \cite{Marklund_PRL_2014, Esirkepov-2015, Kirk-2016} and found to significantly enhance the interaction.

We identify three stages of interaction during the irradiation of the NCD target by four colliding laser pulses. At the first stage, or the initial compression stage, electrons are pushed inward and target is compressed until plasma charge separation field balances the laser ponderomotive force, $F_{p}=\nabla\sqrt{1+a^2}$, where $a=eE_{laser}/m_e\omega $ is the normalized laser amplitude and $\omega$ is the laser frequency. In the following explosion stage, lasers penetrate the NCD plasma and eventually are transmitted due to the self-induced relativistic transparency \cite{cattani2000threshold,tushentsov2001electromagnetic}, which, in principle, can be modified by the QED-effects \cite{1367-2630-17-4-043051}. During this stage energetic electrons collide with laser pulses going through the target (see Fig. \ref{fig1}(b)). After that a two-dimension electromagnetic standing wave(SW) sets up, which traps the electrons at its nodes ($\mathbf{E}=0$ see figure\ref{fig2}(a)) \cite{lehmann2012phase}. These nodes can prevent electrons from leaving the laser volume, leading to overdense or relativistic NCD plasma generation. Finally, there is a saturation stage, when the SW fades away, the production of the $\gamma$ ray photons is terminated and energy fraction of charged particles and $\gamma$ ray photons remains stable. Meanwhile particles escape the confined region and experience a coulomb explosion process. The most important is the second stage, in which brilliant $\gamma$ ray photons are generated.

  \begin{figure}[tbp]
\includegraphics[keepaspectratio=true,height=65mm]
{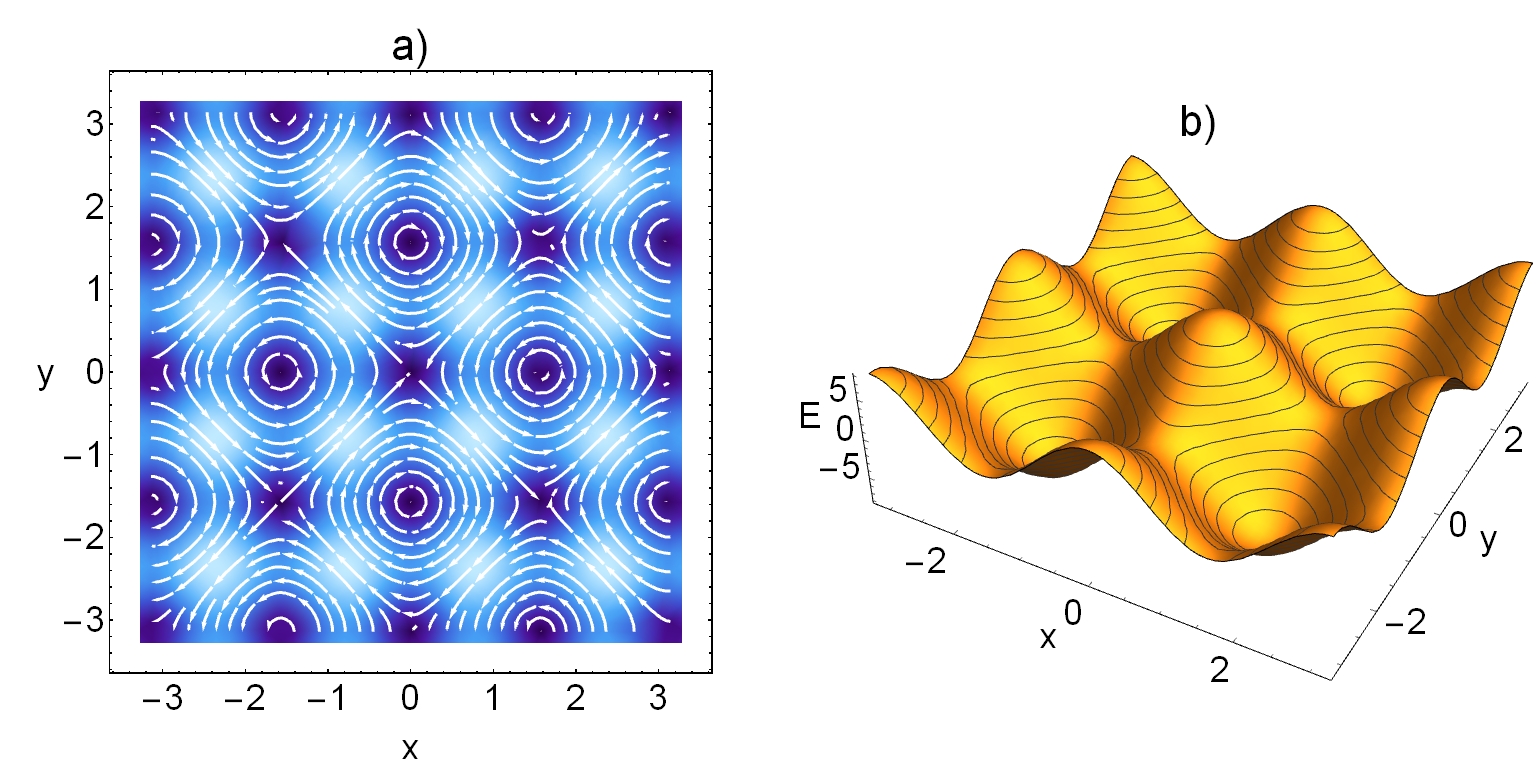}
\caption{
a) Distribution of the magnetic field ${\bf B}(x,y)$ given by Eq. (\ref{eq:Bxy}) in the $(x,y)$ plane; b) $z$-component of the electric field $E(x,y,t)$ given by Eq. (\ref{eq:Ez}) for $\omega=1$, $k=2$, $a=1$ at $t=\pi/4\omega$}.

\label{B-E(x,y)}
\end{figure}

 \begin{figure}[tbp]
\includegraphics[keepaspectratio=true,height=65mm]
{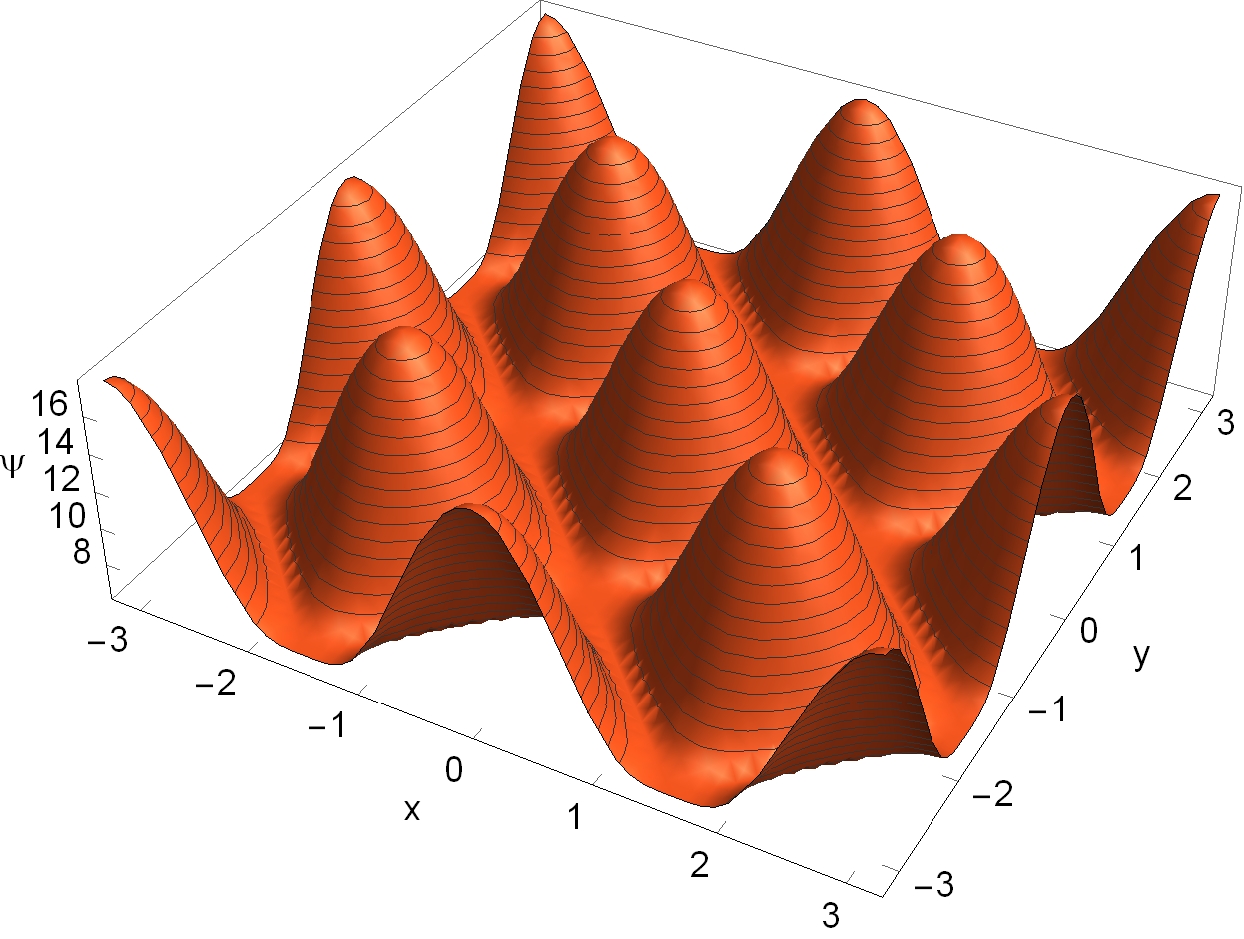}
\caption{
Distribution in the $(x,y)$ plane of the pondemotive potential $\bar{\psi}$ given by Eq. (\ref{eq:psi}).
}
\label{psi}
\end{figure}

\begin{figure}[tbp]
\includegraphics[keepaspectratio=true,height=65mm]
{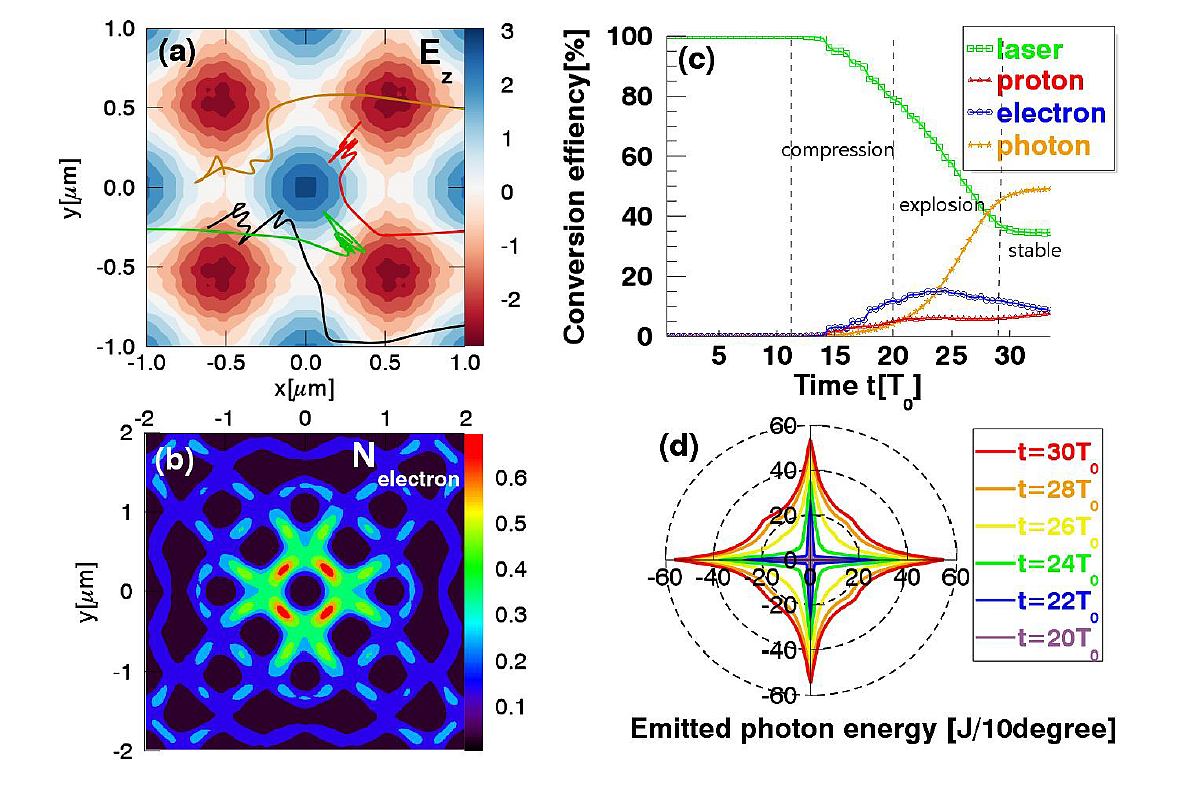}
\caption{
Simulation results under the matching condition ($n_0=16n_c$). (a) Laser amplitude normalized to the maximum of the initial intensity $\rho_l=a_l/a_0$ and trajectories of  test electrons.
 (b) Normalized density of electron $\rho_e=n_e/a_0n_{c}$ at $t=25T_0$.
  (c) Time evolution of the energy of
each of the components (protons, electrons, positrons, photons)as the.  All the energies are normalized to initial total laser energy. (d) Angular distributions of the emitted
 photon energy at different time.
}
\label{fig2}
\end{figure}

The parameter $\chi_e$ for electron in the SW of four laser pulses can be estimated as $\chi_e\approx4\gamma\mathbf{E}/E_{S}$. Depending on the laser intensity we can identify different regimes of interaction. Here we assume that the electron energy is fully determined by its interaction with the EM field. The quantum effect dominated regime is characterized by the fact that the electron is able to emit a photon with the energy of about the electron initial energy. Thus the motion of electrons in this regime is dominated by the quantum recoil. The characteristic field strength in this case is $a_Q=(2\alpha_f/3)^2 (3\lambda/4\pi r_e)$ \cite{Bulanov_PRL_2010_1,Bulanov_NIMA_2011}, where $r_e=2.8\times 10^{-13}$ cm is the classical electron radius and $\alpha_f$ is the fine structure constant, and the condition to be in the quantum dominated regime is $a>a_Q$. Alternatively, if one neglects the actual motion of an electron in strong EM field for simplicity, this condition can be rewritten in the form \cite{di2012extremely}:
\begin{equation}
{\centering \ R_c=\alpha_f\eta a\geqslant1,}
\label{eq1}
\end{equation}
The laser normalized amplitude $a\geqslant175$ can be obtained from above formula in view of $\gamma\approx a$ \cite{meyer2001relativistic}. It should be noted that our regime does not satisfy the requirement of quantum radiation dominated regime \cite{di2012extremely} so both classical and QED electrodynamics analysis can result in the qualitatively identical cooling effect in electron phase space while stochasticity effect in quantum radiation dominated regime \cite{PhysRevLett.111.054802} doesn't play a significant role for the chosen parameters of interaction. Considering the procedure of the compression stage, initial density $n_0$ with size of $S_0$ is gradually piled up to the central area $S_A\approx r_0^2$ with plasma density $n'=n_0S_0/S_A$, where $r_0$ is width of laser beam. On account of relativistic self-induced transparency, the index of refraction is reduced by a factor $1/\sqrt{\bar{\gamma}}$, where $\bar{\gamma}$ is the average Lorentz factor of the electrons. An opaque, or normal over-dense plasma therefore becomes transmissive if the laser amplitude is sufficiently high. Laser energy can be further absorbed in this RCD plasma \cite{freidberg1972resonant,estabrook1975two} and in order to obtain the maximum energy conversion efficiency, initial density can be roughly estimated as

\begin{equation}
{\centering \ n_0S_0/r_0^2\sim\bar{\gamma}n_c,}
\label{eq2}
\end{equation}
$n_c=m_e\varepsilon_0\omega^2/e^2$ is the critical density, where $\varepsilon_0$ is the vacuum permittivity. To achieve high energy conversion, equation (1) and (2) should be satisfied.

In what follows we use EPOCH code \cite{arber2015contemporary}, which is a typical Particle-In-Cell code with a Monte Carlo module, which takes into account the synchrotron emission of $\gamma$ photons and generation of electron-positron pairs \cite{duclous2011monte}. In order to verify this novel regime, QED-PIC simulations are performed. The simulation domain is sampled by $2000\times 2000$ cells,
corresponding to a real space of $40\,\mu m\times40\,\mu m$. Four linearly z-polarized lasers with the same intensities $I_0\approx8.5\times10^{22}\,W/cm^2$
(normalized amplitudes $a_0=eE_L/m_e\omega_0c=250$) and wavelengths $\lambda_0=1.0\,\mu m$ are incident from all sides simultaneously.
The laser pulses have a transverse profile $a\varpropto a_0\exp{(-y^2/r_0^2)}$ of $r_0=4\lambda_0$ and a flat shape duration of $\tau_0=30\,fs(9T_0,T_0=2\pi/\omega_0)$.
The target is a circle located at $x^2+y^2<(8\lambda_0)^2$, with a uniform electron density of $16n_c=1.76\times10^{22}\,cm^{-3}$. Both electrons and protons
with about $6.4\times10^7$ macro-particles are included in this numerical simulation.
Figure \ref{fig2}(a)(b) (case $n_0=16n_c$) shows the overlapped laser field $E_z$ and electron density distribution at $t=25T_0$, where $T_0=3.33fs$ is laser period.
Consistent with the previous theory, electrons are accelerated to hundreds MeV and piled up in central region with area$\sim S_A$ by the surrounding ponderomotive pressure.
At the same time ions are also driven inwards as a result of laser ponderomotive force and the space charge separation field. As shown in figure \ref{fig2}(a),
a two-dimensional SW lattice has been established after counter-propagating lasers penetrate the under-RCD plasma thoroughly. The vector potential of the SW can be taken in the form
\[\mathbf{A}=a_0\left[\cos(\omega t-kx)+\cos(\omega t+kx) +\cos(\omega t-ky)+cos(\omega t+ky)\right ]{\bf e}_z\]
\begin{equation}
\label{eq3}
=2a_0\cos \omega t \left(\cos kx+\cos ky \right) {\bf e}_z.
\end{equation}
From Eq. (\ref{eq3}) it follows that electric and magnetic field can be given as
\begin{equation}
\mathbf{E}=-\partial_t\mathbf{A}=2a_0 \omega \sin \omega t \left(\cos kx+\cos ky \right) {\bf e}_z
\label{eq:Ez}
\end{equation}
 and
\begin{equation}
\mathbf{B}=\nabla\times\mathbf{A}=-2 a_0 k \cos \omega t \left(\sin ky \, {\bf e}_x-\sin kx  \, {\bf e}_y \right) ,
\label{eq:Bxy}
\end{equation}
 respectively.

The SW node ($E_z=0$) locates at $x+y=(n+1/2)\lambda_0$ or $x-y=(n+1/2)\lambda_0$ (see fig \ref{E(x,y)}). Under such circumstance, the ponderomotive force
exerts on electrons in $(x,y)$ plane is from $\nabla\psi$, here ponderomotive potential $\psi(x,y,t)=\sqrt{1+|{\bf A}(x,y,t)|^2}$ and its period average is given as
\[
\bar{\psi}=\frac{\omega}{2\pi}\int_{-\pi/\omega}^{\pi/\omega}\psi(x,y,t)dt=\frac{\omega}{\pi} \left[E\left(-4 a^2 (\cos (k x)+\cos (k y))^2\right)\right.
\]
\begin{equation}
\left .+\sqrt{4 a^2 (\cos (k x)+\cos (k y))^2+1}\, E\left(\frac{4 a^2 (\cos (k x)+\cos (k y))^2}{4 a^2 (\cos (k x)+\cos (k y))^2+1}\right)\right],
\label{eq:psi}
\end{equation}
where $E(x)$ is the complete elliptical integral of the second kind.
The potential $\bar{\psi}$ shown in Fig. \ref{psi} demonstrates a tendency of electrons to move along the valleys.

The numerical result of $\bar{\psi}$ at $t=25T_0$ is in figure\ref{fig2}(c) which indicates its valley has the identical location with electric node ($x+y=(n+1/2)\lambda_0$ or $x-y=(n+1/2)\lambda_0$). In figure\ref{fig2}(b), electron lattice with normalized density  $\rho_e=n_e/a_0n_c\lesssim1$, transparent to lasers $(a_0=250)$, is built and the electron average density increasing from initial $16n_c$ up to $100n_c$ with an area of $S_A\sim (1.4r_0)^2\approx30\mu m^2$ approximately in agreement with eq.2. Obviously electron spatial distribution figure2(b) is in agreement with $\bar{\psi}$ valley in Fig.\,2(c). The distribution of dimensionless relativistic invariant parameters $\eta$ at $t=25T_0$ is plotted in figure\ref{fig2}(d) which illustrates a large fraction of electron with $\eta\sim0.1$ so that in our condition considerable photons are emitted while there is no occurrence of the quantum stochasticity inducing the spread of phase space \cite{PhysRevLett.111.054802}.

The evolution of the particle and laser energy fraction is in Fig.\,\ref{fig2}(e). Obviously the $\gamma$ photon generation mainly emerges from $20T_0$ to $29T_0$.
Actually the explosion stage can be subdivided into two phases. The first is collision at $20T_0<t<24T_0$ where electrons are mainly backscattered
with the laser pulses and radiate many energetic $\gamma$ photons along their velocity direction. In Fig.\,\ref{fig2}(f), the production of backscattering photons is dominant at angle $\theta=90^\circ,180^\circ,270^\circ$ and $360^\circ$ during $20T_0<t<24T_0$, where $\theta$ is photon momentum direction in $X-Y$ plane. The second is MCS where the electron lattice has been constructed in central region during $24T_0<t<29T_0$. Because of relativistic effects, electron can be released from the ponderomotive trapping and move chaotically as the field amplitude rises \cite{bauer1995relativistic}. As the role of radiation losses increases, following "phase space contraction" \cite{tamburini2011radiation}, the particles subsequently become trapped once more and again in the electric field nodes \cite{lehmann2012phase}, which explains the electron spatial distribution in Fig.\,\ref{fig2}(b). During this phase, electrons don't have preponderant kinetic direction and photons are emitted isotropically. Some energetic electron trajectories are depicted in Fig.\,\ref{fig2}(a), showing they may oscillate at particular angle $\theta=45^\circ,135^\circ,225^\circ,315^\circ$ when confined in the electric field node. Consequently in Fig.\,\ref{fig2}(f) a tiny peak emerges at $\theta=45^\circ,135^\circ,255^\circ,315^\circ$. At $t=25T_0$ photon density at central area can attain $10^{30}/m^3$ and positron density maximum are larger than $10^{27}/m^3$. After the interaction, the number of $\gamma$ ray photons is about $6.1\times10^{14}$ and positrons is $1.9\times10^{10}$ respectively obtaining $47\%$ and $0.03\%$ of the total laser energy. The number of $\gamma$ photon produced per electron per laser period is given by $N_\gamma\approx6.42\alpha_f\bar{\gamma}$  \cite{bell2008possibility}. For four lasers of intensity $9\times 10^{22}\,W/cm^2$ focused onto a NCD target, $N_\gamma\approx5.91\times 10^{14}$ which are agreeable with our simulation result. Average photon energy is $10.8\,MeV$ and parameter $\chi\approx\bar{\gamma}E/E_S\approx10^{-3}$ by simulation. The total pair-production rate per electron per laser period is the product of the photon absorption probability $1-e^{-\tau}$ in a length $\lambda_{laser}$ multiplied by the rate of production of photons by curvature radiation \cite{bell2008possibility}. The photon optical depth to absorption in a path length $\lambda_{laser}$ is $\tau\approx12.8(I_{24}\lambda_{\mu m}^2)e^{[-4/(3\chi)]}\approx0$ \cite{ridgers2013dense}, for $\chi\ll1$ and hence most of gamma-photon energy can be preserved.

\begin{figure}[tbp]
\includegraphics[keepaspectratio=true,height=35mm]{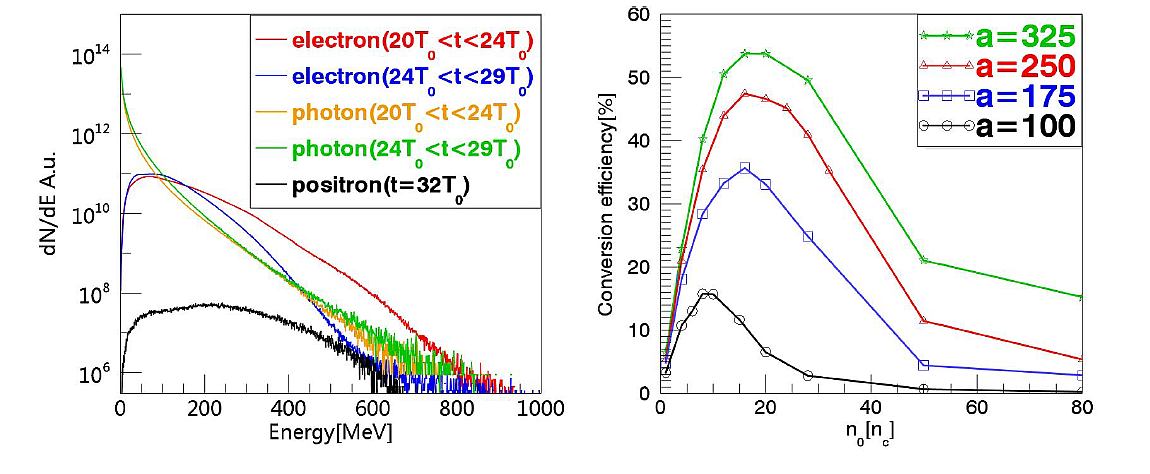}
\caption{Energy spectrum and conversion efficiency: (a) Energy spectra of initial density($n_0=16n_c$).
Red (blue) line represents average electron spectrum at $20T_0<t<24T_0$ ($24T_0<t<29T_0$).
 Orange (green) lines depicts spectra of produced photon at $20T_0<t<24T_0$ ($24T_0<t<29T_0$).
 (b) Conversion efficiency from laser to $\gamma$-ray photons scaling with the initial target density($n_0$) in our scheme.}
\label{fig3}
\end{figure}

Fig.\,\ref{fig3}(a) exhibits the energy spectrum in the $n_0=16n_c$ case. Energetic electrons are drastically decelerated after colliding with laser on the opposite side. Both the backscattering ($20T_0<t<24T_0$) and MCS ($24T_0<t<29T_0$) phases in explosion stage produce considerable $\gamma$ ray photons which is corresponding with the electron energy evolution in figure\ref{fig2}. It shows MCS plays a significant role in conversion efficiency of gamma photons. At $t=33T_0$, average energy of electron and positron are 90\,MeV and 236\,MeV. Since the positron is likely to be produced in more intense SW field, it can be immediately accelerated after generation. Conversion efficiency from lasers to photons versus diverse target density is displayed in Fig.\,\ref{fig3}(b). Due to the QED nonlinear effect \cite{di2012extremely}, production of $\gamma$ ray photons growth nonlinearly as the laser intensity increases.  Eq.\,(1) amounts the approximative threshold of ($a_{thr}\sim175$) where nonlinear QED effects is dominated in laser electron interaction. At $a_0>a_{thr}$, radiation recoil is so tremendous that electrons cannot sustain high energy when experiencing the SW optical lattice. The averaged electron energy is approximately $\bar{\gamma_e}mc^2\approx100\,MeV$ and the optimistic initial density can be estimated as $n_{opt}\sim\bar{\gamma_e}n_cS_A/S_0\approx16 n_c(\bar{\gamma_e}\approx200)$. With laser amplitude $a=325$, the $\gamma$ photon conversion efficiency attains 53.7\% and the average electron energy is just 101\,MeV when optimistic condition $n_0=16n_c$ and Eqs. (1,2) are satisfied, which is in good agreement with above theoretical model.

\begin{figure}[tbp]
\includegraphics[keepaspectratio=true,height=65mm]
{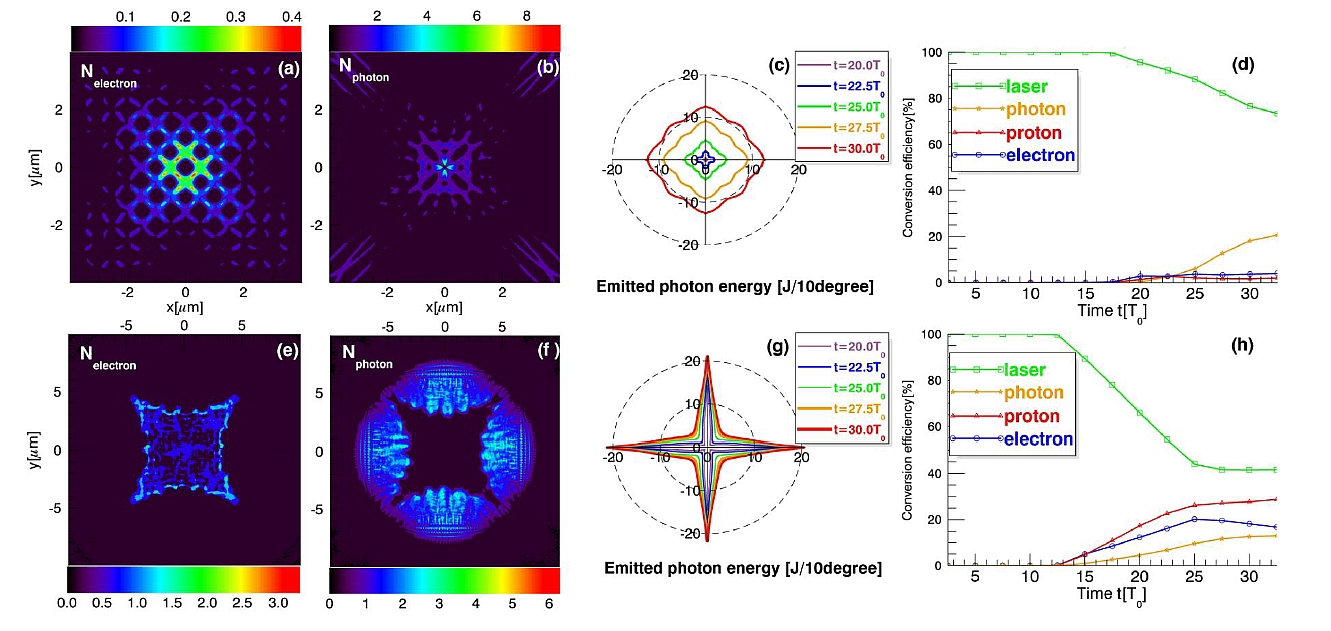}
\caption{Simulations in mismatching condition ($n_0=4n_c$) and ($n_0=50n_c$):
(a) and (b) show the normalized density of electrons and radiation photons at $t=25T_0$, where the initial density is chosen to be $n_0=4n_{c}$,
breaching the matching condition (3).
(c) is the angular distributions of the emitted photon energies and (h) shows the conversion efficiency as function of time. (e)(f)(g)(h) are the same as above, where initial density is $n_0=50n_{c}$.}
\label{fig4}
\end{figure}

If the optimal matching condition (Eqs.(1,2)) is not satisfied, the energy conversion efficiency will be reduced.
Figs.\,\ref{fig4}(a--d)  show the results of simulations for $n_0=4n_{c}$, where target is so dilute that the lasers easily penetrate it without much energy transferred to plasma and gamma photon. The biggest disadvantage of $n_0=4n_{c}$ case is lack in electrons. Occurrence of the collision process ($20T_0<t<24T_0$) is reduced so drastically that most of photons is radiated at MCS ($24T_0<t<29T_0$) in fig\ref{fig4} (d) and the angular distribution of photon energy is nearly isotropic in Fig.\,\ref{fig4} (c). At $t=33T_0$ $\gamma$ photon number with the average energy $13.7MeV$ is $2\times 10^{14}$, nearly one-third of the above matching condition $n_0=16n_{c}$. Only $20\%$ energy is transferred from laser to photons.
Figs.\,\ref{fig4} (e--h) present the overdense target $n_0=50n_{cr}$ case, where intense pulses are mainly reflected at the over RCD interface and the SW electron lattice can not be formed. This circumstance is similar with relativistic hole boring regime\cite{naumova2009hole, SlowWave}, when the laser radiation is reflected from the overdense plasmas in Fig.\,\ref{fig4} (e). The radiation front moves with the relativistic velocity $v_{HB}$ equal to $v_{HB}/c=\sqrt{\Xi}/(1+\sqrt{\Xi})$, where $\xi=I/\rho c^3$ is the dimensionless intensity of the laser piston. Under the simulation conditions, the velocity of the interface and accelerated protons energy can be estimated as $v_{HB}\approx0.4c$ and $2\xi m_ic^2/(1+2\sqrt{\xi})\approx350MeV$, respectively, which agrees with simulation results presented  in Fig.\,\ref{fig4}(f). The emitted photons propagate predominantly along the lasers propagation direction as seen in Fig.\,\ref{fig4}(g). The energy conversion efficiency is about  $10\%$, while a considerable part of the laser energy is transferred to protons (see Fig. \ref{fig4}(h)).

In conclusion, a novel laser plasma based gamma-ray source is proposed and investigated systematically with PIC simulations. By irradiating NCD targets with four symmetrical imploding pulses, bright gamma-rays can be generated by MCS and nearly half of the laser energy is transferred to the gamma-rays. With a simple model, the matching conditions to achieve high conversion rate and the optimal target density are given. It should be noted that our regime is still valid in 3D circumstance. Utilizing the same parameters as 2D optimal condition in figure2, we can get the identic electron optical lattice distribution (cross section in $z=0$ plane) and $36\%$ laser energy converted to gamma-rays. The only difference between 3D and 2D condition is the emergency of the electron dispersing on Z-direction due to the transverse ponderomotive force. Therefore with powerful laser facilities such as ELI under construction, such a gamma-ray source is promising to be realized in the near future.

The work has been supported by the National Basic Research Program of China (Grant No.2013CBA01502), NSFC (Grant Nos.11535001) and National Grand Instrument Project (2012YQ030142). The PIC code Epoch is funded by the UK EPSRC grants EP/G054950/1, EP/G056803/1, EP/G055165/1 and EP/ M022463/1. The PIC simulations were carried out in Super Computation Center in Max Planck Institute and Shanghai Super Computation Center. The author Z. Gong acknowledges useful discussion with Dr. M. L. Zhou at LMU. Z. Gong and R. H. Hu contributed equally to this work.

\bibliography{aa}



\end{document}